\documentclass[twocolumn,aps,superscriptaddress,showpacs,nofootinbib,floatfix]{revtex4}
\usepackage{listings}
\usepackage[utf8]{inputenc}
\usepackage{hyperref}
\usepackage{amsmath}
\usepackage{amssymb}
\usepackage{amsfonts}
\usepackage{tabularx}
\usepackage{booktabs}
\usepackage{graphicx}
\usepackage{color}
\usepackage{multirow}
\usepackage[inline]{enumitem}
\usepackage[T2A,T1]{fontenc}
\graphicspath{{fig/}}
\definecolor{theblue}{RGB}{0,50,230}

\usepackage{hyperref}
\hypersetup{
  colorlinks=true,
  linkcolor=theblue,
  citecolor=theblue,
  urlcolor=theblue
}

\newcommand{\sqrts}{\sqrt{s_{NN}}}

\newcommand {\avg}[1]{\ensuremath{\langle\kern-1.0pt\langle#1\rangle\kern-1.0pt\rangle}}

\newlength\cmsFigWidth

\usepackage[normalem]{ulem}  % \sout{old text} for strikeout

\renewcommand\sout{\bgroup \color{red} \ULdepth=-.5ex \ULset}
\begin{document}

\title{Spectra and flow of light nuclei in relativistic heavy ion collisions at RHIC and the LHC}

\author{Wenbin Zhao}
\affiliation{Department of Physics and State Key Laboratory of Nuclear Physics and Technology, Peking University, Beijing 100871, China}
\affiliation{Collaborative Innovation Center of Quantum Matter, Beijing 100871, China}
\author{Lilin Zhu}
\affiliation{Department of Physics, Sichuan University, Chengdu 610064, China}
\author{Hua Zheng}
\affiliation{School of Physics and Information Technology, Shaanxi Normal University, Xi'an 710119, China}
\affiliation{Laboratori Nazionali del Sud, INFN, via Santa Sofia, 62, 95123 Catania, Italy}
\author{Che Ming Ko}
\affiliation{Cyclotron Institute and Department of Physics and Astronomy, Texas A$\&$M University, College Station, TX 77843, USA}
\author{Huichao Song}
\affiliation{Department of Physics and State Key Laboratory of Nuclear Physics and Technology, Peking University, Beijing 100871, China}
\affiliation{Collaborative Innovation Center of Quantum Matter, Beijing 100871, China}
\affiliation{Center for High Energy Physics, Peking University, Beijing 100871, China}

\date{\today}

\begin{abstract}
Within the framework of the coalescence model based on the phase-space distributions of protons and neutrons
generated from the {{\tt iEBE-VISHNU}} hybrid model with {{\tt AMPT}} initial conditions, we study the spectra and elliptic flow of deuterons and helium-3 in relativistic heavy ion collisions at the Relativistic Heavy Ion Collider (RHIC) and the Larger Hadron Collider (LHC). Results from our model calculations for Au + Au collisions at $\sqrt{s_{NN}}=200$ GeV at RHIC and Pb+Pb collisions at $\sqrt{s_{NN}}=2.76$ TeV at the LHC are compared with available experimental data. Good agreements are generally seen between theoretical results and experimental data, except that  the calculated yield of helium-3 in Pb + Pb collisions at $\sqrt{s_{NN}}=2.76$ TeV underestimates the data by about a factor of two.  Possible reasons for these discrepancies are discussed. We also make predictions on the spectra and elliptic flow of deuterons and helium-3 in Pb + Pb collisions at $\sqrt{s_{NN}}=5.02$ TeV that are being studied at LHC.
\end{abstract}
\maketitle

%After produced at chemical freeze-out surface through the statistical hadronization mechanism,
%this extended blast-wave model takes into consideration of the larger in-plane than out-of-plane flow velocity due to the positive elliptic flow by introducing a space-momentum correlation in the nucleon phase-space distribution.

%\blue{It was found that\cite{Oh:2009gx} the hadronic transport model including the production and annihilation of deuteron and  the coalescence model give the very  similar  $p_T$-spectra of deuteron, while the elliptic flow obtained in the two model are quite different  at Au + Au 200 GeV.}
%And at LHC collision energy, ALICE Collaboration also measured production of  light -nuclei in Pb + Pb at 2.76 TeV \cite{Acharya:2017dmc}.

%they could be break apart during hadronic evolution and then regenerated by the final-state coalescence mechanism.

%((Besides fixing its parameters by fitting the measured proton transverse momentum spectrum and elliptic flow,
%The extended blast-wave model was motivated by its similarity to the freeze-out configuration of a real dynamical model, but it is not really true that the hydrodynamical freeze-out configuration corresponds to the parameter set that test describes the experimental data.

%Furthermore, with ten freely tunable parameters for the phase space distribution of freeze-out nucleons, the extended blast-wave model is clearly a toy model with little predictive power. ))

\section{Introduction}

The production of light nuclei and their antiparticles in relativistic heavy ion collisions is a topic of great interest as it provides a unique opportunity to investigate whether nuclei and antinuclei have the same properties and to discover if exotic antinuclei can exist in nature~\cite{Abelev:2009ae,Agakishiev:2011ib,Sharma:2011ya,Adam:2015vda}. Recently, the production of light nuclei has been widely measured at RHIC at BNL~\cite{Adler:2001uy,Abelev:2010rv,Agakishiev:2011ib,Yu:2017bxv} and the LHC at CERN~\cite{Adam:2015yta,Adam:2015vda,Acharya:2017dmc}. The related theoretical investigations have been carried out in the frameworks of either the statistical model or the coalescence model. In the statistical model, the yields of hadrons and light nuclei can be nicely described with a few parameters related to the chemical freeze-out conditions~\cite{Andronic:2010qu,Cleymans:2011pe}. In the coalescence model, light nuclei are formed through the recombination of protons and neutrons with close positions and velocities on the kinetic freeze-out surface \cite{Chen:2012us,Chen:2013oba,Shah:2015oha,Sun:2015jta,Sun:2015ulc,Botvina:2017yqz,Yin:2017qhg,Zhu:2017zlb,Oh:2009gx}.

%at RHIC collision energy, And  at LHC collision energy,
Although both the statistical model and the coalescence model can describe the yields of light nuclei at RHIC and  LHC, it is generally believed that light nuclei formed from the statistical hadronization at chemical freeze-out can hardly remain stable during the hadronic evolution due to their small binding energies. Recently, the STAR, PHENIX and ALICE Collaborations have further measured the elliptic flow of deuterons ($d$) and helium-3 ($^3$He) in Au+Au collisions at $\sqrt{s_{NN}}=200$ GeV~\cite{Adler:2004uy,Abelev:2009ae,Adamczyk:2015ukd,Adamczyk:2016gfs} and in Pb+Pb collisions at $\sqrt{s_{NN}}=2.76$ TeV~\cite{Adam:2015vda,Acharya:2017dmc}. It was found that the elliptic flow of these two light nuclei could not be described with the blast-wave model through simply replacing the proton mass with the ones of the light nuclei~\cite{Adamczyk:2016gfs}. Also, the simplest or naive coalescence model~\cite{Kolb:2004gi}, which assumes that the deuteron elliptic flow at certain transverse momentum is twice of that of the proton at half of the transverse momentum~\cite{Acharya:2017dmc}, failed to reproduce the measured elliptic flow of deuterons. On the other hand, Refs.~\cite{Yin:2017qhg,Zhu:2017zlb} showed that the spectra and elliptic flow of light nuclei could be described by the coalescence model using the phase-space distribution of nucleons generated from a modified blast-wave model.  Besides the usual parameters, such as the fugacity $\xi$, the kinetic freeze-out temperature $T_K$, the nucleon emission proper time $\tau_0$, the radial flow velocity $\beta(r)$ and $p_T$-dependent elliptic flow harmonic coefficient $\varepsilon(p_T^{})$, etc., which are required to fit the spectra and elliptic flow of pions, kaons and protons, additional space-momentum correlations between protons and neutrons are required in this study to fit the elliptic flow of deuterons and helium-3.

These previous studies~\cite{Yin:2017qhg,Zhu:2017zlb} have revealed that the formation of light nuclei is very sensitive to the phase-space distributions of protons and neutrons at kinetic freeze-out, which are strongly influenced by the dynamical evolution of the QGP and the hadronic fireball.  In the standard model of relativistic heavy ion collisions, the evolving system after thermalization is described by hydrodynamics for the QGP fluid followed by a hadron cascade simulations for the hadronic evolution~\cite{Song:2010aq,Heinz:2013th,Gale:2013da,Song:2013gia,Song:2017wtw,Schenke:2010rr,Shen:2014vra}. One such standard model is the {{\tt iEBE-VISHNU}} hybrid model~\cite{Shen:2014vra} that combines the (2+1)-d viscous hydrodynamics~\cite{Song:2007fn,Song:2009gc} with the UrQMD hadron cascade model~\cite{Bass:1998ca,Bleicher:1999xi}. During the past few years, {{\tt iEBE-VISHNU}} has been widely used to study various soft physics at RHIC and LHC, and this has led to many nice descriptions and predictions of various flow data, including flow harmonics, flow distributions, symmetric cumulants, event plane correlations, etc.~\cite{Song:2013qma,Zhu:2015dfa,Xu:2016hmp,Qian:2016pau,Zhu:2016puf,Zhao:2017yhj,Zhao:2017rgg,Song:2017wtw}. In particular, the {{\tt iEBE-VISHNU}} hybrid model with {\tt AMPT} initial conditions has successfully described and predicted the spectra and flow harmonics of pions, kaons and protons in Pb+Pb collisions at $\sqrt{s_{NN}}=2.76$ and 5.02 TeV~\cite{Xu:2016hmp,Zhao:2017yhj,Adam:2016nfo,Acharya:2018zuq}. The {{\tt iEBE-VISHNU}}+UrQMD model with  {\tt AMPT} initial conditions  is thus expected to provide the realistic phase-space distributions of protons and neutrons that are needed for the coalescence calculations of  light nuclei production in relativistic heavy ion collisions. In this paper, we will use the nucleon phase-space distributions from this model to study and predict the spectra and elliptic flow of deuterons and helium-3 in Au+Au collisions at $\sqrt{s_{NN}}=200$ GeV and in Pb+Pb collisions at $\sqrt{s_{NN}}=2.76$ and 5.02 TeV in the framework of coalescence model.  We emphasize that the parameters in {{\tt iEBE-VISHNU}} simulations have been fixed by the yields, spectra and flow harmonics of all charged hadrons in previous studies~\cite{Xu:2016hmp,Zhao:2017yhj}.  Compared with the modified blast-wave model~\cite{Yin:2017qhg,Zhu:2017zlb} that introduces a parametrized space-momentum correlation of nucleons to describe the elliptic flow of light nuclei, such space-momentum correlations of emitted hadrons are naturally included in {{\tt iEBE-VISHNU}} through the dynamically generated nucleon distribution function without additional free parameters~\cite{Song:2010aq}.

This paper is organized as the following: Sec.~II  and Sec~III briefly introduce the coalescence model, the {\tt iEBE-VISHNU} hybrid model  and the set-ups of the calculations.  Sec.~IV presents and discusses the spectra and elliptic flow of deuterons and helium-3 in Au+Au collisions at $\sqrt{s_{NN}}=200$ GeV and Pb+Pb collisions at $\sqrt{s_{NN}}=2.76$ and 5.02 TeV, calculated with the coalescence model using phase-space distributions of nucleons from {\tt VISHNU}. Sec.~V summarizes the results and gives the conclusion from the present study.

\section{The coalescence model}\label{coalescence}
In the coalescence model~\cite{Mattiello:1996gq,Chen:2003qj,Chen:2003ava}, the production probability of a nucleus of atomic number $A$ is given by the overlap of the Wigner function  $f_A({\bf x}_1^\prime, ... ,{\bf x}_A^\prime; {\bf p}_1^\prime, ... ,{\bf p}_A^\prime, t^\prime)$ of the nucleus with the phase-space distributions $f({\bf x}_i, {\bf p}_i, t)$
of nucleons at the kinetic freeze-out:
\begin{eqnarray}
\label{coal}
&&\frac{d^3N_A}{d{\mathbf P}_A^3}= g_A\int \Pi_{i=1}^Ap_i^\mu d^3{\sigma_{i\mu}} \frac{d^3{\bf p}_i}{E_i}
f({\bf x}_i, {\bf p}_i, t)\nonumber\\
&&\times f_A({\bf x}_1^\prime, ... ,{\bf x}_A^\prime; {\bf p}_1^\prime, ... ,{\bf p}_A^\prime; t^\prime)\delta^{(3)}\left({\bf P}_A-\sum_{i=1}^A{\bf p}_i\right),\nonumber\\
\end{eqnarray}
where $g_A=(2J_A+1)/[\Pi_{i=1}^{A}(2J_i+1)]$ is the statistical factor for $A$ nucleons of spins $J_i$ to form a nucleus of angular momentum $J_A$. The coordinate and momentum of the $i$-th nucleon in the fireball frame are denoted by ${\bf x}_i$ and ${\bf p}_i$, respectively. Their coordinate ${\bf x}_i^\prime$ and momentum ${\bf p}_i^\prime$ in the Wigner function of the produced nucleus are obtained by Lorentz transforming the coordinate ${\bf x}_i$ and momentum ${\bf p}_i$ to the rest frame of the nucleus and then propagating earlier freeze-out nucleons freely with a constant velocity, determined by the ratio of its momentum and energy in the rest frame of the nucleus, to the time $t^\prime$ when the last constituent nucleon in the nucleus freezes out.

In this paper, we focus on investigating deuteron and helium-3 production and elliptic flow. Following Ref. \cite{Chen:2003ava}, their corresponding Wigner functions are obtained from the Wigner transformation of their wave functions, which are taken to be the product of the harmonic oscillator wave functions.  In this case, the Wigner function of deuteron is~\cite{Song:2012cd}
\begin{eqnarray}
f_2(\boldsymbol\rho,{\bf p}_\rho)=8g_2\exp\left[-\frac{\boldsymbol\rho^2}{\sigma_\rho^2}-{\bf p}_\rho^2\sigma_\rho^2\right],
\label{two}
\end{eqnarray}
where the relative coordinate $\boldsymbol\rho$ and relative momentum ${\bf p}_\rho$ are defined as:
\begin{eqnarray}\label{rel}
\boldsymbol\rho=\frac{1}{\sqrt{2}}({\bf x}_1^\prime-{\bf x}_2^\prime),\quad{\bf p}_\rho=\sqrt{2}~\frac{m_2{\bf p}_1^\prime-m_1{\bf p}_2^\prime}{m_1+m_2},
\end{eqnarray}
with $m_i$ being the mass of nucleon $i$. The width parameter $\sigma_\rho$ in Eq. (\ref{two}) is related to the root-mean-square radius of  deuteron via~\cite{Song:2012cd}
\begin{eqnarray}
\langle r_{d}^2 \rangle=\frac{3}{2}\frac{m_1^2+m_2^2}{(m_1+m_2)^2} \sigma_\rho^2=\frac{3}{4}\frac{m_1^2+m_2^2}{\omega  m_1m_2(m_1+m_2)},
\end{eqnarray}
if we use the relation $\sigma_\rho=1/\sqrt{\mu_1\omega}$ in terms of the oscillator frequency $\omega$ in the harmonic wave function and the reduced mass $\mu_1=2(1/m_1+1/m_2)^{-1}$.

Similarly, the Wigner function of helium-3 is~\cite{Song:2012cd}
\begin{eqnarray}
&&f_3(\boldsymbol\rho,\boldsymbol\lambda,{\bf p}_\rho,{\bf p}_\lambda)\nonumber\\
&&=8^2g_3\exp\left[-\frac{\boldsymbol\rho^2}{\sigma_\rho^2}-\frac{\boldsymbol\lambda^2}{\sigma_\lambda^2}-{\bf p}_\rho^2\sigma_\rho^2-{\bf p}_\lambda^2\sigma_\lambda^2\right],
\label{three}
\end{eqnarray}
where $\boldsymbol\rho$ and ${\bf p}_\rho$ are similarly defined as in Eq. (\ref{rel}), and the relative coordinate $\boldsymbol\lambda$ and momentum ${\bf p}_\lambda$ are defined as:
\begin{eqnarray}
{\boldsymbol\lambda}&=&\sqrt{\frac{2}{3}}\left(\frac{m_1{\bf x}_1^\prime+m_2{\bf x}_2^\prime}{m_1+m_2}-{\bf x}_3^\prime\right),\nonumber\\
{\bf p}_\lambda&=&\sqrt{\frac{3}{2}}~\frac{m_3({\bf p}_1^\prime+{\bf p}_2^\prime)-(m_1+m_2){\bf p}_3^\prime}{m_1+m_2+m_3}.
\end{eqnarray}
%and the width parameter $\sigma_\rho$ is the same as in Eq. (\ref{two}).
The width parameter $\sigma_\lambda$ in Eq.(\ref{three}) is related to the oscillator frequency by $\sigma_\lambda=1/\sqrt{\mu_2 \omega}$ with $\mu_2=(3/2)[1/(m_1+m_2)+1/m_3]^{-1}$.  The values of $\sigma_{\rho}$ and $\sigma_\lambda$ are  determined from the oscillator constant via the root-mean-square radius of helium-3 by ~\cite{Song:2012cd}
%\begin{widetext}
\begin{eqnarray}
&&\langle r_{He}^2 \rangle=\nonumber\\
&&\frac{1}{2}\frac{m_1^2(m_2+m_3)+m_2^2(m_3+m_1)+m_3^2(m_1+m_2)}{\omega(m_1+m_2+m_3)m_1m_2m_3}.\nonumber\\
\end{eqnarray}
%\end{widetext}

For the production of helium-3, besides the coalescence process $p+p+n\rightarrow^3$He from two protons and one neutron, we also include the coalescence process $d+p\rightarrow ^3$He from  one deuteron and one proton. Here, the deuteron is treated as a point particle with its phase-space distribution given by that obtained from the coalescence of proton and neutron, the corresponding two-body Wigner function of helium-3 is given by Eq. (\ref{two}) with $\boldsymbol\rho$ and ${\bf p}_\rho$ denoting the relative coordinate and momentum between deuteron and proton.

The statistical factors and  values of the width parameters in the Wigner functions for deuteron and helium-3 as well as the empirical values of their charge radii and  resulting oscillator constants are given in TABLE \ref{tab}.

\begin{table}[h]
\caption{{\protect\small Statistical factor ($g$), radius ($R$), oscillator frequency ($\omega$), and width parameters ($\sigma_\rho$, $\sigma_\lambda$) for deuteron and helium-3. Radii are taken from Ref.~\cite{Rappold:2013jta}}. }
\label{tab}
\begin{tabular}{c|cccccc}
\hline\hline
Nucleus & $g$ & R (fm) & $\omega$ (sec$^{-1})$ & $\sigma_\rho, \sigma_\lambda$ (fm) \\
\hline
deuteron & 3/4 & 2.1421 & 0.1739 & 2.473 \\
$p+p+n\rightarrow^3$He & 1/4 & 1.9661 & 0.5504 & 1.390 \\
$d+p\rightarrow^3$He & 1/3 & 1.9661 & 0.3389 & 1.536 \\
\hline\hline
\end{tabular}
\end{table}

%\section{{\tt iEBE-VISHNU} hybrid model and set-ups}
\section{The {\tt iEBE-VISHNU} hybrid model}
\label{sec:model}

In this paper, we use the {\tt iEBE-VISHNU} hybrid model to generate the phase-space distributions of nucleons for the 
coalescence model calculations of light nuclei.  The {\tt iEBE-VISHNU} model~\cite{Shen:2014vra} is an event-by-event  version of the {\tt VISHNU} hybrid model~\cite{Song:2010aq} that combines the (2+1)-d viscous hydrodynamics {\tt VISH2+1}~\cite{Song:2007fn,Song:2009gc} for the QGP expansion with the hadron cascade model UrQMD~\cite{Bass:1998ca,Bleicher:1999xi} for the evolution of subsequent hadronic matter. In the present study, we follow Refs.~\cite{Zhao:2017yhj,Xu:2016hmp} to set the starting time of hydrodynamics at $\tau_0=0.6\ \mathrm{fm}/c$ and implement the {{\tt AMPT}} initial conditions to generate the fluctuating initial profiles of energy density in the transverse plane with zero transverse velocity.  Also, the normalization factors for the initial energy density profiles are tuned to fit the yields of all charged hadrons in Au + Au collisions at $\sqrt{s_{NN}}=200$ GeV or in Pb + Pb collisions at $\sqrt{s_{NN}}=2.76$ and 5.02 TeV. In {\tt iEBE-VISHNU}, the transport equations for energy-momentum tensor $T^{\mu \nu}$ and the 2nd order Israel-Stewart equations for shear stress tensor $\pi^{\mu \nu}$ and bulk pressure $\Pi$ for the QGP evolution are solved by using the (2+1)-d viscous hydrodynamics {\tt VISH2+1} with  longitudinal boost invariance~\cite{Song:2007fn,Song:2007ux,Song:2009gc}.  Following~Refs. \cite{Zhao:2017yhj,Xu:2016hmp}, we set the specific shear viscosity and specific bulk viscosity in QGP to $\eta/s=0.08$ and $\zeta/s=0$, respectively, and use the equation of state from the HotQCD Collaboration~\cite{Bernhard:2016tnd,Moreland:2015dvc}. The switching temperature $T_{sw}$ that controls the transition from the hydrodynamic evolution to the hadron cascade simulation is set to 148 MeV.  After the UrQMD hadronic evolution, we output the coordinate and momentum information of various stable hadrons at the kinetic freeze-out, which can be used to calculate the spectra and flow observables as well as to provide the phase-space distributions of nucleons for the coalescence model study of light nuclei production.

We emphasize that with the above set-ups and the fine tuned parameters in {\tt AMPT} initial conditions ~\cite{Zhao:2017yhj,Xu:2016hmp}, {\tt iEBE-VISHNU} could nicely describe various soft hadron data at RHIC and LHC, especially on the $p_T$ spectra and flow harmonics of pions, kaons and protons ~\cite{Zhao:2017yhj,Xu:2016hmp}. This indicates that {\tt iEBE-VISHNU} can generate proper phase-space distributions of nucleons at the kinetic freeze-out for the coalescence model calculations.  Compared with previous studies based on the blast-wave model~\cite{Yin:2017qhg,Zhu:2017zlb}, the space-momentum correlations of nucleons are naturally included in {{\tt iEBE-VISHNU}} through the dynamically generated distribution functions without introducing any free parameters~\cite{Song:2010aq}. Details of parameter fitting and results on the spectra and various flow data of all charged and identified hadrons within the framework of {\tt iEBE-VISHNU} can be found in Refs.~\cite{Zhao:2017yhj,Xu:2016hmp}.

\section{RESULTS}\label{sec:results}

%We simulate the heavy-ion collisions at RHIC and LHC using {{\tt iEBE-VISHNU}} hybrid model with {\tt{AMPT}} initial condition, adopting the parameters given in section \ref{sec:model}, until the system evolves to kinetic freeze-out respectively. The phase space distributions of protons and neutrons needed in the coalescence model can be obtained.

%The good agreements with experimental data, both for the particle spectra and elliptic flow have been found. We also make the predictions for deuteron in Pb + Pb collisions at $\sqrts=$ 5.02 TeV.

Using the phase-space distributions of protons and neutrons from the {\tt iEBE-VISHNU} hybrid model with {\tt AMPT} initial conditions  as described in Section~\ref{sec:model}, we show in the following the transverse momentum spectra and elliptic flow of deuterons and helium-3 in Au + Au collisions at $\sqrt{s_{NN}}=200$ GeV and in Pb + Pb collisions at $\sqrt{s_{NN}}=2.76$ and 5.02 TeV obtained from the coalescence model described in Section~\ref{coalescence}.

%the  with {\tt AMPT} initial conditions can describe not only  the flow and spectra of hadrons, but also the $v_n$ distribution,  the mode coupling effects and non-linear response of QGP evolution within one set parameters. }

%For $p_T > $ 3.2 GeV, where hydrodynamics is not supposed to work properly because the hard processes are dominant in heavy-ion collisions, such as jet and/or minijets, the hydro calculations slightly underestimate the proton spectrum.

%is the transverse momentum spectrum of deuteron from the coalescence model using the phase-space distributions of protons and neutrons from {{\tt iEBE-VISHNU}} hybrid model,

\subsection{Au + Au collisions at $\sqrt{s_{NN}}=$ 200 GeV}

Figure~1 shows the transverse momentum spectra of protons, deuterons and helium-3 in Au + Au collisions at $\sqrt{s_{NN}}=$ 200 GeV and 0-80\% centrality. The spectrum of protons is calculated from the {{\tt iEBE-VISHNU}} hybrid model with {\tt AMPT} initial conditions , which nicely describes the PHENIX data below 3 GeV.  With the phase-space distributions of protons and neutrons from {{\tt iEBE-VISHNU}}, we calculate the spectra of deuterons and helium-3 using the coalescence model. As shown by the blue dotted line in Fig.~1, our model calculation nicely reproduces the deuteron data from PHENIX~\cite{Adler:2004uy}. For the transverse momentum spectrum of helium-3, our calculations that include the two coalescence processes $p+p+n\rightarrow^3$He and $p+d\rightarrow^3$He roughly describe the data from STAR, while including only the single coalescence process  $p+p+n\rightarrow^3$He significantly under-predicts the data  for helium-3 as described in details in the Appendix. We thus include in the present study both coalescence processes to calculate the spectrum and elliptic flow of helium-3 as in Ref.~\cite{Zhang:2018euf}. 

%The possible coalescence channel of \red{di-protons} with a neutron is not considered because the bound state of two protons has not been found so far.
%After taking into account these two above coalescence processes, the spectrum of helium-3 is also roughly reproduced.

%
\begin{figure}[pht]
  \centering \includegraphics[scale=0.4]{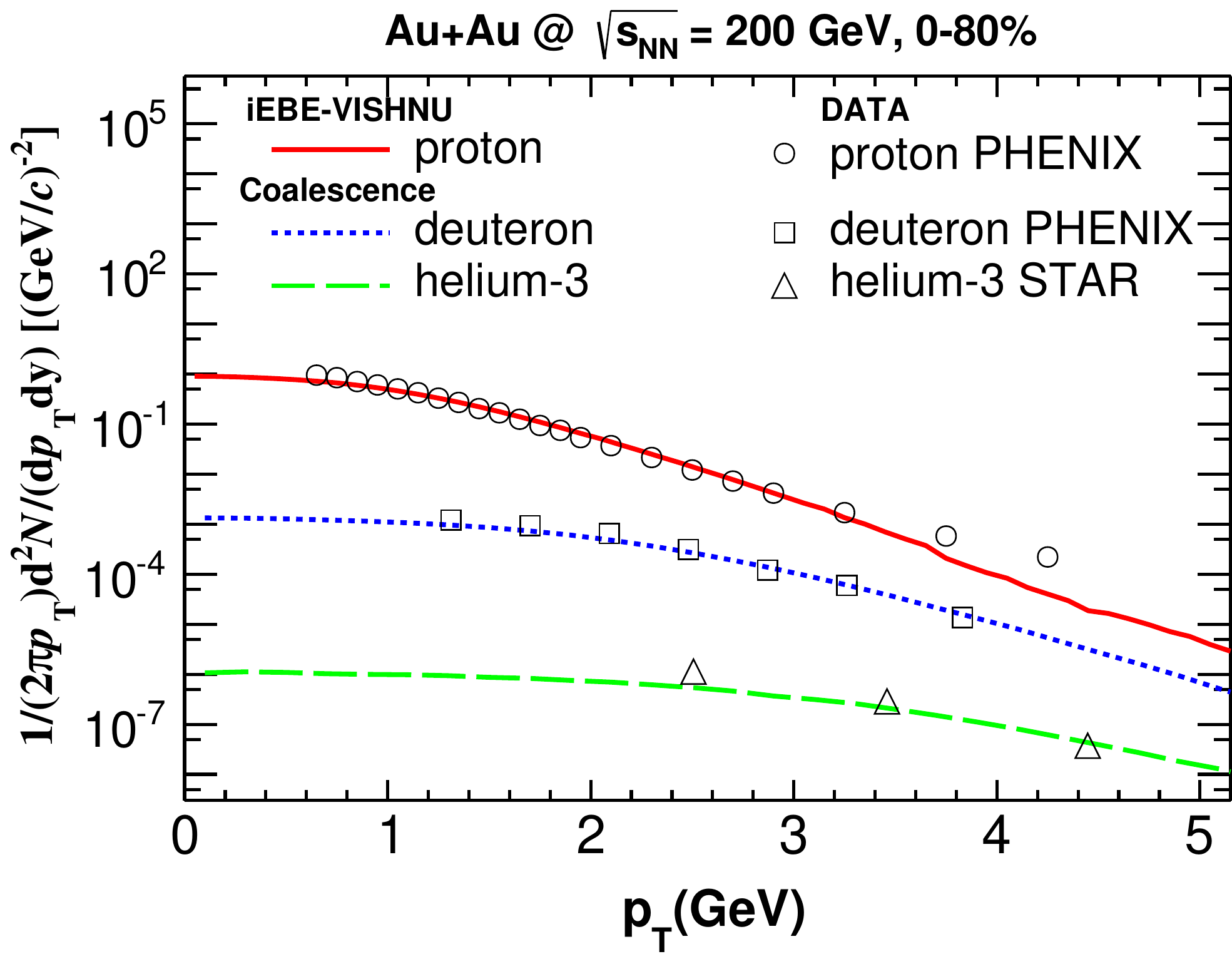}
  \caption{Transverse momentum spectra of protons, deuterons and helium-3 in Au + Au collisions at $\sqrt{s_{NN}}=$ 200 GeV and 0-80\% centrality, calculated from the {{\tt iEBE-VISHNU}} hybrid model (protons) and  from the coalescence model (deuterons and helium-3) using the phase-space distributions of protons and neutrons from {{\tt iEBE-VISHNU}}.  The data for protons, deuterons and helium-3 are taken from the PHENIX~\cite{Adler:2003cb,Adler:2004uy} and STAR~\cite{Abelev:2009ae} Collaborations, respectively. }
  \label{auauspectra}
\end{figure}
\begin{figure}[pht]
  \centering \includegraphics[scale=0.4]{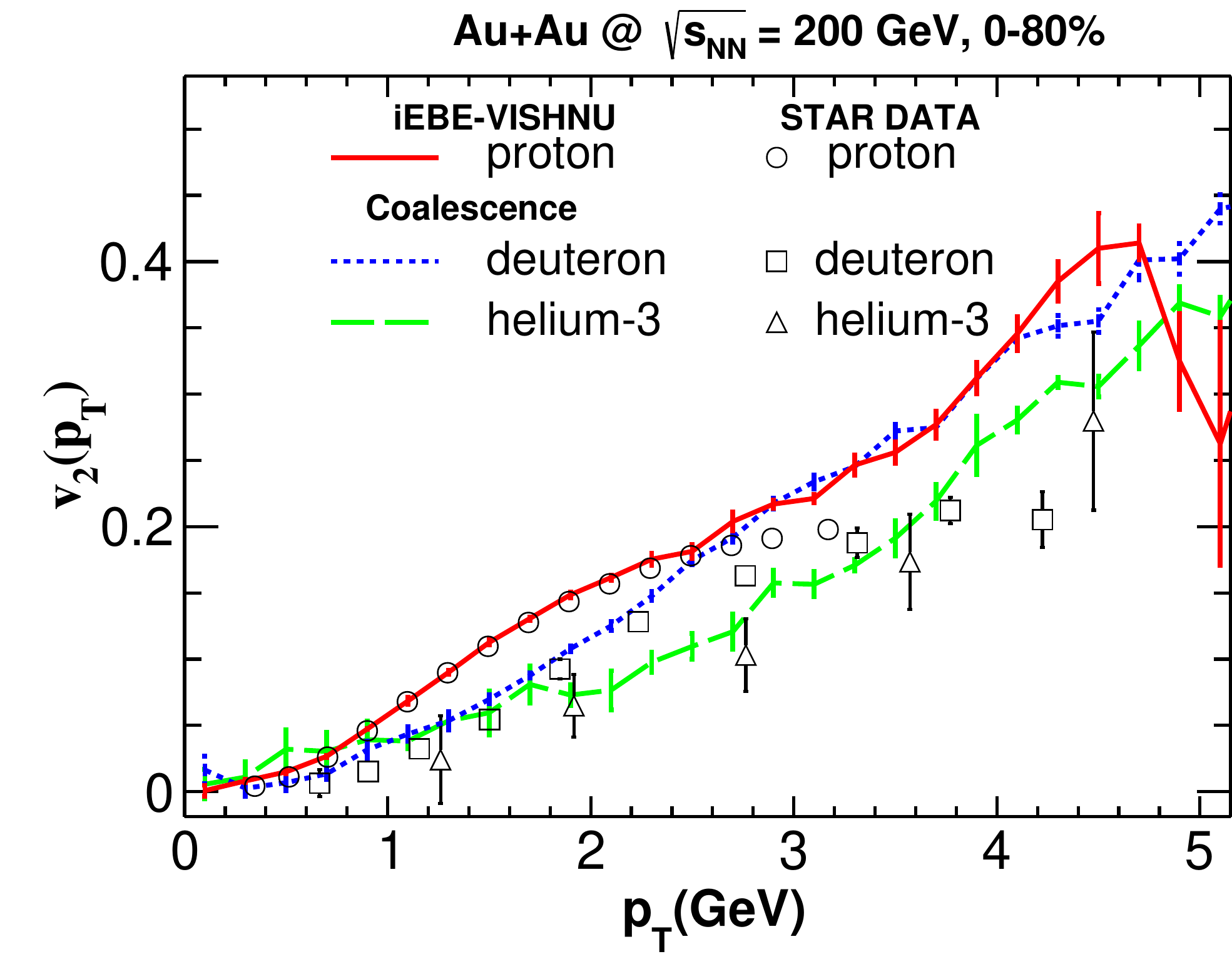}
  \caption{Elliptic flow of protons, deuterons and helium-3 in  Au + Au collisions at $\sqrt{s_{NN}}=$ 200 GeV and 0-80\% centrality, calculated from {{\tt iEBE-VISHNU}} (protons) and  from the coalescence model (deuterons and helium-3).  The data of protons, deuterons and helium-3 are taken from the STAR Collaboration~\cite{Adamczyk:2015ukd,Adamczyk:2016gfs}.}
  \label{auauv2pt}
\end{figure}

\begin{figure*}[t]
  \includegraphics[scale=0.72]{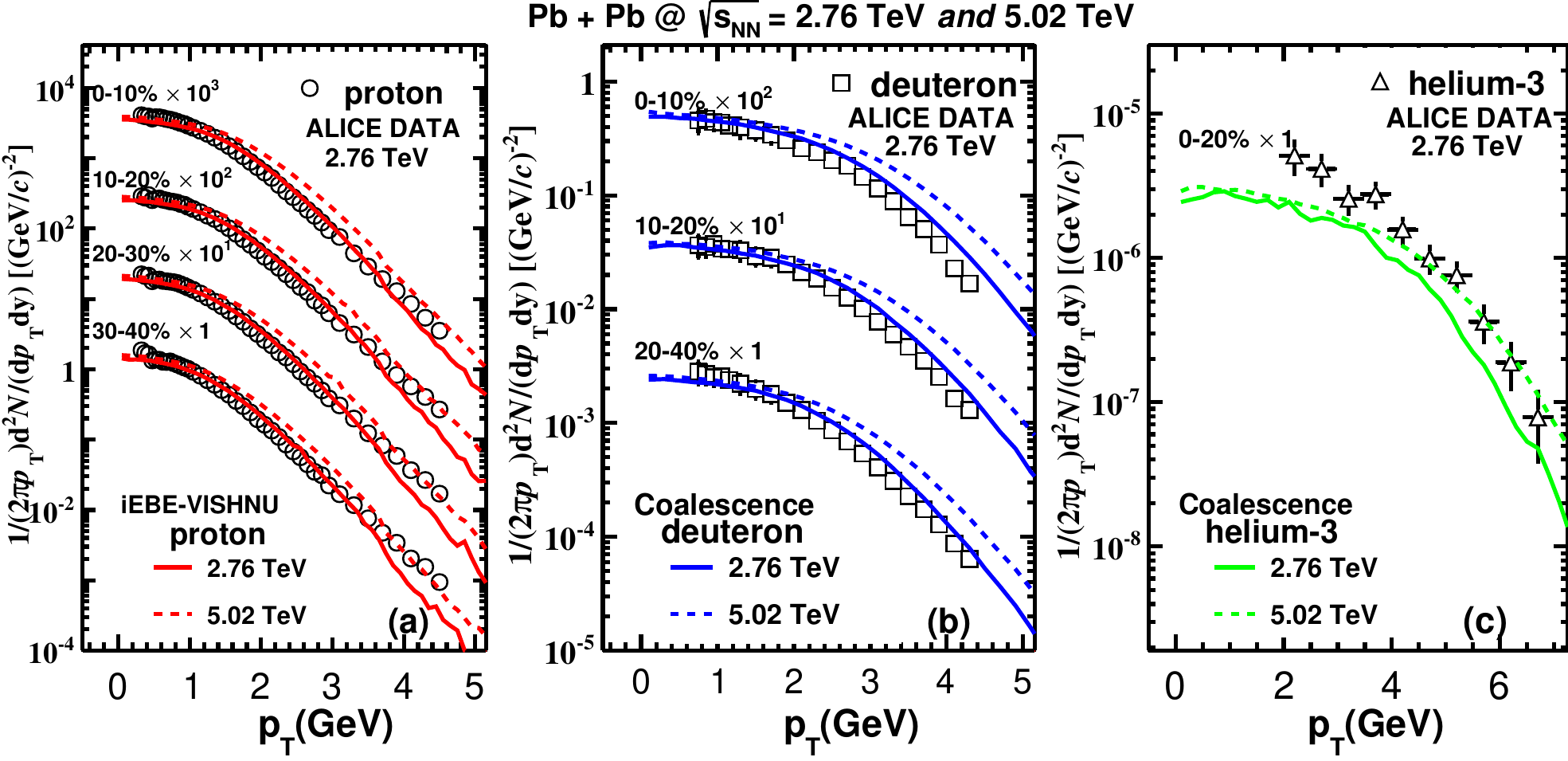}
  \caption{Transverse momentum spectra of protons, deuterons and  helium-3 in Pb + Pb collisions at $\sqrts=$ 2.76 and 5.02 TeV, calculated from {\tt iEBE-VISHNU} (protons) and from the coalescence model (deuterons and helium-3). The data for protons, deuterons and helium-3 at $\sqrts=$ 2.76 and 5.02 TeV are taken from Refs.~\cite{Abelev:2013vea,Acharya:2017dmc} and \cite{Adam:2015vda}, respectively.}
  \label{pbpb_spectra}
\end{figure*}

\begin{figure*}[t]
  \includegraphics[scale=0.9]{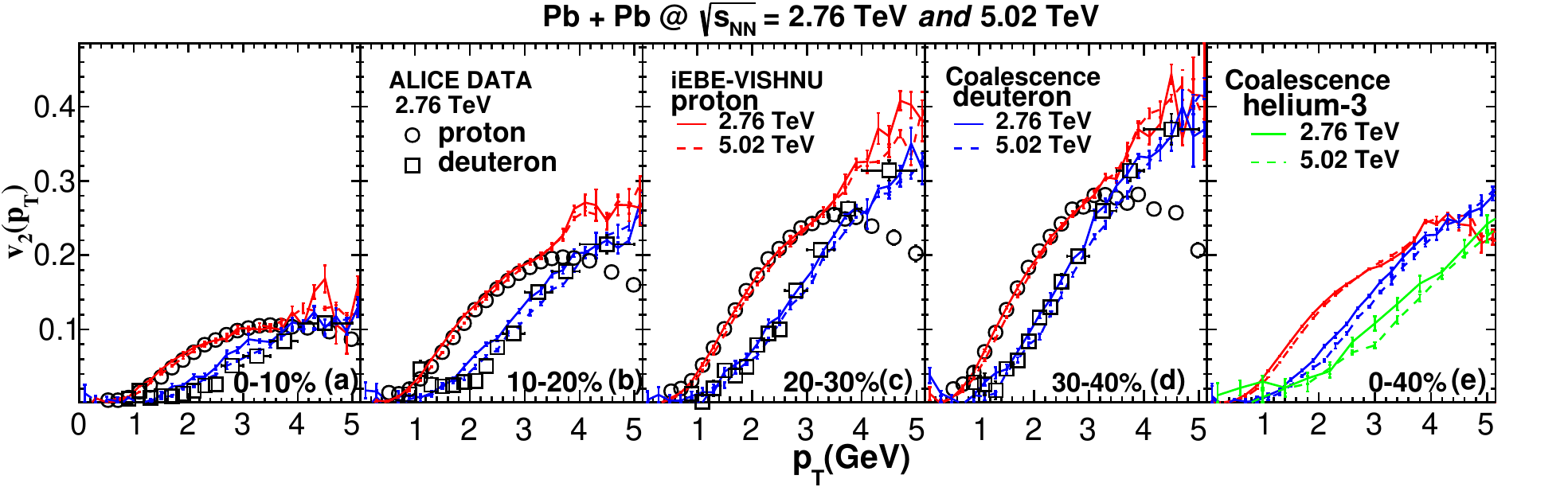}
  \caption{Elliptic flow of protons, deuterons and helium-3 in Pb + Pb collisions at $\sqrts=$ 2.76 and 5.02 TeV, calculated from {{\tt iEBE-VISHNU}} (protons) and  from the coalescence model (deuterons and helium-3). The data for protons and deuterons  at  $\sqrts=$ 2.76 TeV  are taken from Refs.~\cite{Adam:2016nfo} and \cite{Acharya:2017dmc}, respectively.  }
  \label{pbpb_v2pt}
\end{figure*}

Figure 2 shows the differential elliptic flow $v_2(p_T)$ of protons, deuterons and helium-3 in Au + Au collisions at $\sqrt{s_{NN}}=$ 200 GeV and 0-80\% centrality. Following Refs.~\cite{Adamczyk:2015ukd,Adamczyk:2016gfs}, we calculate $v_2(p_T)$ using the event-plane method with particle of interest (POIs) and reference particles (RPs) selected from two sub-events with their pseudorapidity $\eta$ taken within $-1.0<\eta<0$ and $0<\eta<1.0$, respectively. It is seen that the {{\tt iEBE-VISHNU}} hybrid model with {\tt AMPT} initial conditions  and other tuned parameters~\cite{Xu:2016hmp,Zhao:2017rgg} gives an overall quantitative description of $v_2(p_T)$ of protons below 2.5 GeV but becomes increasingly larger than the data as the proton momentum increases, where the hydrodynamic approach becomes not applicable.  The elliptic flow of deuterons and helium-3 from the coalescence model using the phase-space distributions of protons and neutrons from {{\tt iEBE-VISHNU}} can roughly describe the measured data from the STAR Collaboration, including the clear $v_2$ mass ordering among protons, deuterons and helium-3. On the other hand,  the measured and calculated elliptic flow of deuterons gradually deviate from each other above 2 GeV in Au+Au collisions at $\sqrt{s_{NN}}=$ 200 GeV, which is surprising as one would expect a well-tuned phase-space distributions of protons below 2 GeV be able to describe the elliptic flow of deuterons up to a momentum of 4 GeV. This discrepancy between the model result and  data from 2 to 4 GeV for deuterons at RHIC requires further study.

%Also shown in the figure are the proton (open circles), deuteron (open squares)  and helium-3 (open triangle) transverse momentum spectra from ALICE Collaboration \cite{Abelev:2013vea, Adam:2015vda, Acharya:2017dmc}.

\subsection{Pb + Pb collisions at $\sqrts=$ 2.76 and 5.02 TeV}

Figure~3 shows the transverse momentum spectra of protons, deuterons and helium-3 in Pb + Pb collisions at $\sqrts=$ 2.76 and 5.02 TeV. Similar to Fig.~1 for Au+Au collisions at $\sqrt{s_{NN}}=$ 200 GeV,  the spectra of protons shown in the left panel of Fig. 3 are calculated from {{\tt iEBE-VISHNU}} with {{\tt AMPT}} initial conditions  nicely describes the ALICE data below 3 GeV in Pb+Pb collisions at  $\sqrt{s_{NN}}=2.76$ TeV for various centralities. With re-tuned parameters that fit the measured multiplicity~\cite{Zhao:2017rgg}, we predict the spectra of protons in Pb + Pb collisions at $\sqrt{s_{NN}}=$ 5.02 TeV with {{\tt iEBE-VISHNU}}, which are seen to be higher and flatter than the ones at $\sqrt{s_{NN}}=$ 2.76 TeV due to larger multiplicity and stronger radial flow~\cite{Zhao:2017yhj}. The spectra of deuterons obtained with the coalescence model using the phase-space distributions of protons and neutrons from {\tt iEBE-VISHNU} are shown in the middle panel of Fig.~3. Our calculations are seen to describe the low-$p_T$ spectra of deuterons from ALICE for Pb + Pb collisions at $\sqrt{s_{NN}}=$ 2.76 TeV and  0-10\%, 10-20\% and 20-40\% centralities, but to slightly over-predict the data above 2.5 GeV for these three centralities. We also predict the transverse momentum spectra of deuterons in Pb + Pb collisions  at $\sqrt{s_{NN}}=$ 5.02 TeV, which are higher and flatter than the ones at $\sqrt{s_{NN}}=$ 2.76 TeV as the case for protons. The right panel of Fig. 3  shows the transverse momentum spectrum of helium-3 in Pb +Pb collisions at $\sqrt{s_{NN}}=$ 2.76 and 5.02 TeV for 0-20\% centrality that are calculated and predicted from the coalescence model respectively. As in the calculations for Au+Au collisions at $\sqrt{s_{NN}}=$ 200 GeV, we include both coalescence processes $p+p+n\rightarrow^3$He and $d+p\rightarrow^3$He, which approximately double the yield of helium-3 for the case of the single coalescence process $p+p+n\rightarrow^3$He. However, the yield of helium-3 in Pb +Pb collisions  at  $\sqrt{s_{NN}}=$ 2.76 TeV is still largely underestimated, which also leads to an under-prediction of its spectrum. Detailed discussions on this are given in the Appendix.

%\footnote{The differential elliptic  flow harmonics of helium-3 can also be predicted by {{\tt iEBE-VISHNU}} + coalescence model. However, it requires much higher statistical runs to reduce the error bars. Therefore, we don't further predict them here. }.

%for the centralities of $0-10\%$, $10-20\%$, $20-30\%$ and $30 -40\%$

%Again, the theoretical results of protons are from {\tt iEBE-VISHNU} hybrid model, while the ones of deuteron and helium-3 are from the coalescence model.

Figure~4 shows the elliptic flow of  protons, deuterons and helium-3 in Pb + Pb collisions at   $\sqrt{s_{NN}}=$ 2.76 and 5.02 TeV. Following Refs.~\cite{Adam:2016nfo,Acharya:2017dmc}, we use the Q-cumulant method to calculate $v_2(p_T)$ for protons and deuterons. Shown in the four panels to the left of Fig. 4 are the results at $0-10\%$, $10-20\%$, $20-30\%$, and $30 -40\%$ centralities, respectively. Results from our calculations give a nice description of the measured $v_2(p_T)$ of protons up to 3 GeV at these centralities in Pb + Pb collisions at $\sqrt{s_{NN}}=$ 2.76 TeV~\cite{Acharya:2018zuq}. The red dashed lines are predicted $v_2(p_T)$ of protons at 5.02 TeV, which are slightly lower than the ones at 2.76 TeV due to the slightly increased radial flow at larger collision energy, and this is consistent with recent ALICE experimental data~\cite{Acharya:2018zuq}. For the deuteron elliptic flow in collisions at 20-30\% and 30-40\% centralities, our model calculations based on the coalescence model using the phase-space distributions of protons and neutrons from {{\tt iEBE-VISHNU}} can quantitatively describe its momentum dependence up to 4 GeV within error bars. However, for collisions at $0-10\%$ and $10-20\%$ centralities, our model calculations slightly over-predict the data above 1.5 GeV, although the input phase-space distribution of protons describes the corresponding $v_2(p_T)$ of protons well. The predicted  elliptic flow of deuterons at $\sqrts$= 5.02 TeV are slightly below the one at 2.76 TeV, which is similar to the case of protons.

Due to limited statistics, we only calculate the elliptic flow of helium-3 at $0-40\%$ centrality as shown in 
the right panel of Fig. 4.  As in recent experimental measurements~\cite{QM2018:Puccio}, we use the Event Plane (EP) method with two sub-events within $-1.0<\eta<0$ and $0<\eta<1.0$ to calculate the $v_2(p_T)$ of protons, deuterons and helium-3 at  this centrality  instead of the Q-cumulant method used for other centralities.  Results from our model calculations show a clear $v_2$ mass ordering among protons, deuterons and helium-3. With increasing collision energies, the  difference between protons and helium-3 $v_2$ also increases. Such $v_2$ mass ordering is similar to that in  Au+Au collisions at $\sqrts=$ 200 GeV, which can also be tested in future experiments.

% . Recently, ALICE Collaboration measured the anisotropic flow of inclusive charged particles in Pb+Pb collisions at $\sqrts=$ 2.76 TeV and 5.02 TeV \cite{Acharya:2018lmh}. The experimental data show that the difference of the elliptic flow between the two collision energies is indeed very small, which is consistent with our results in the present study.

%It is should be noted that the spatial and momentum distributions of protons and neutrons are from  {\tt{iEBE-VISHNU}} hybrid model, which means hydrodynamics + UrQMD generates proper space-momentum correlations of nucleons at kinetic freeze-out proposed in Ref. \cite{Yin:2017qhg}.

%\subsection{$p_T$-spectra and $v_2(p_T)$ of helium-3 calculated by  iEBE-VISHNU + coalescence model}
%\subsection{Remarks on helium-3 (I modify here. Maybe we should move this part before the discussion of Fig. 4.) }

%\com{Delete the following paragraph}
%Figure ~ \ref{auauspectra} and panel (c) in Fig. ~\ref{pbpb_spectra} show the $p_T$-spectra of helium-3 in $0-80\%$ centrality bin at Au + Au 200 GeV and 0-20\% at Pb +Pb 2.76 TeV respectively. The generations of helium-3 at Au + Au can be approximately described by {{\tt iEBE-VISHNU}} + coalescence model of the summation over $p+p+n$ and $p+d$ two channels. However, at LHC collision energy,  coalescence model significantly  underpredict helium-3.  (Similar results can be seen at Ref.\cite{Yin:2017qhg,Zhu:2017zlb}.)
%This gives us the confidence to claim that

\section{Summary}\label{sec:summary}
In this paper, we have studied   the spectra and elliptic flow of deuterons and helium-3 in
relativistic heavy ion collisions at RHIC and LHC in the coalescence model.  The needed phase-space distributions of protons and neutrons at kinetic freeze-out are generated from the {{\tt iEBE-VISHNU}} hybrid model
with  {{\tt AMPT}} initial conditions, which have been shown to describe nicely the spectra and
elliptic flow of pions, kaons and protons in previous calculations~\cite{Zhao:2017yhj,Xu:2016hmp}. Results from our coalescence model calculations are found to roughly describe the $p_T$ spectra of deuterons and the differential elliptic flow of deuterons and helium-3 at various centralities in Au + Au collisions at $\sqrt{s_{NN}}=$ 200 GeV and in Pb + Pb collisions at $\sqrt{s_{NN}}=$ 2.76 TeV.  We have also predicted the spectra and elliptic flow of deuterons and helium-3 in Pb + Pb collisions at  $\sqrt{s_{NN}}=$ 5.02 TeV, which can be compared with experimental measurements in the near future. The agreement between our model calculations and  available data at RHIC and LHC indicates that the coalescence model, together with the proper phase-space distributions of nucleons generated from  a realistic hybrid model,  supports the picture that light nuclei are produced through the coalescence of protons and neutrons in relativistic heavy ion collisions at RHIC and LHC.

Although the coalescence model nicely describes the elliptic flow of helium-3 at RHIC and the LHC,  there exists a discrepancy between the calculated and measured $p_T$ spectrum of helium-3 in Pb + Pb collisions at $\sqrt{s_{NN}}=$ 2.76 TeV.  Although including the two coalescence processes $p+p+n\rightarrow^3$He and $p+d\rightarrow^3$He in our coalescence model calculations can roughly describe the $p_T$ spectra of helium-3 in  Au + Au collisions at $\sqrt{s_{NN}}=$ 200 GeV, it under-predicts the yield of helium-3 in Pb + Pb collisions at $\sqrt{s_{NN}}=$ 2.76 TeV. The reason for this discrepancy may be due to  its earlier formation than the deuteron in collisions at higher energies as discussed in the Appendix.

%We also predicted the $p_T$-spectra and  differential elliptic flow of  deuterons and helium-3  in 5.02 A TeV Pb + Pb collisions from the coalescence model. The spectra of  protons, deuterons and helium-3 at $\sqrt{s_{NN}}=$ 5.02 TeV are flatter, and elliptic flow of protons, deuterons and helium-3 at $\sqrt{s_{NN}}=$ 5.02 TeV are  slightly lower than the ones at 2.76 TeV due to the slightly increased radial flow at larger collision energy.
%We hope this theoretical predictions  could provide  some suggestions for the  coming experimental mesurements.

\section*{Acknowledgements}
We thank the discussions from K.~Sun, H.~Xu and J.~Zhao.  W.~Z. and H.~S. are supported by the NSFC and the MOST under grant Nos. 11435001, 11675004 and 2015CB856900. L.~L.~Z is supported in  part by the NSFC of China under Grant no. 11205106. C.~M.~K is supported by the US Department of Energy under Contract No. DE-SC0015266 and the Welch Foundation under Grant No. A-1358. W.~Z. and H.~S. also gratefully acknowledge the extensive computing resources provided by the Super-computing Center of Chinese Academy of Science (SCCAS), Tianhe-1A from the National Supercomputing Center in Tianjin, China and the High-performance Computing Platform of Peking University.

%\section*{Appendix: the production of helium-3}
\section*{Appendix: helium-3 production}
\label{sec:appendix}

In this appendix, we discuss in detail the production of helium-3 in relativistic heavy ion collisions at RHIC and LHC in the framework of the coalescence model that uses the phase-space distribution of kinetically freeze-out nucleons from the {{\tt iEBE-VISHNU}} model with {\tt AMPT} initial conditions. As briefly mentioned in Sec.~IV, the yield of helium-3 measured in experiments cannot be reproduced by the coalescence model that includes only the  coalescence process $p+p+n\rightarrow^3$He. A similar result has  been found in a recent work on hypertriton production in relativistic heavy ion collisions~\cite{Zhang:2018euf}. In that study, the experimental data cannot be explained by the  coalescence process $p+n+\Lambda\rightarrow^3_{\Lambda}$H alone. After also including the process $d+\Lambda\rightarrow^3_{\Lambda}$H, the yield of $^3_{\Lambda}$H is found to be enhanced by about a factor of two, which helps to achieve a better description of the experimental data. Following Ref.~\cite{Zhang:2018euf}, we include in our study both processes $p+p+n\rightarrow^3$He and $d+p\rightarrow^3$He in the coalescence calculations of helium-3.

%To illustrate the effects from different coalescence processes,

\begin{figure*}[t]
  \includegraphics[scale=0.7]{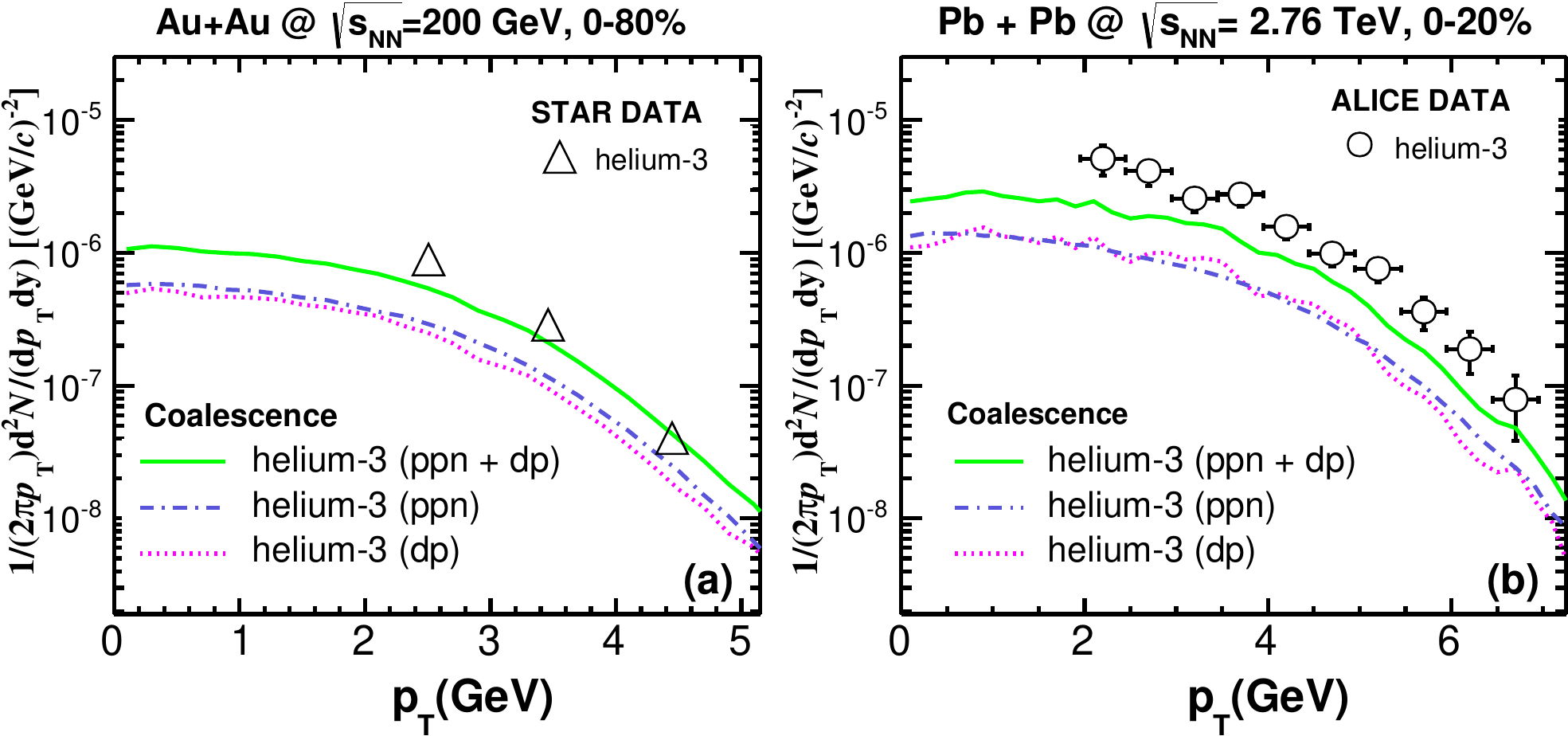}
  \caption   {Transverse momentum spectra of helium-3 in Au + Au collisions at $\sqrt{s_{NN}}=$ 200 GeV and 0-80\% centrality and in Pb + Pb collisions at $\sqrt{s_{NN}}=$ 2.76 TeV and 0-20\% centrality, calculated from the coalescence model using the phase-space distributions of protons and neutrons from {{\tt iEBE-VISHNU}}. The blue dashed-dotted line and the pink dotted line are from the coalescence processes of $p+p+n\rightarrow^3$He and $p+d\rightarrow^3$He, respectively. The green solid line is the sum of these two processes. The data in Au + Au and Pb + Pb collisions are taken from the STAR Collaboration~\cite{Abelev:2009ae} and the ALICE Collaboration~\cite{Adam:2015vda}, respectively.  }
  \label{helium3_spectra}
\end{figure*}

In Fig.~5, we show the $p_T$-spectra of helium-3 in Au + Au collisions at $\sqrt{s_{NN}}=$ 200 GeV and in Pb + Pb collisions at $\sqrt{s_{NN}}=$ 2.76 TeV, calculated from the coalescence model with $p+p+n\rightarrow^3$He (blue dashed-dotted line), with $d+p\rightarrow^3$He (pink dotted line) and with both these two processes (green solid line). At both RHIC and LHC collision energies, the spectra of helium-3 from the two coalescence processes $p+p+n\rightarrow^3$He and $p+d\rightarrow^3$He are close to each other. Adding the two contributions thus enhances the yield of helium-3 by about a factor of two, which roughly describes the measured $p_T$ spectrum  in Au+Au collisions. However,  the calculated spectrum  of helium-3 in Pb+Pb collisions from  both coalescence processes still under-predicts the data.

% \red{They  found the constancy of the deuteron number} during hadronic evolution in the transport model}

In Ref.~\cite{Oh:2009gx}, deuteron production in Au + Au collisions at $\sqrt{s_{NN}}=$ 200 GeV has been studied within the transport approach by assuming that deuterons are initially statistically produced at  chemical freeze-out and then followed by a hadronic evolution that includes both their annihilations and reproductions via the reactions $NN\leftrightarrow d\pi$. It is found that such transport approach gives a similar $p_T$ spectrum of deuterons as the one obtained from the coalescence model using the phase-space distributions of nucleons at kinetic freeze-out . This study thus indicates that for deuteron production at top RHIC energy, the coalescence model based on  protons and neutrons at the kinetic freeze-out is a good approximation to the transport approach that takes into account their annihilations and reproductions.  However, the production of helium-3 has not been investigated within the transport approach since it is highly nontrivial to include the three-particle interaction for helium-3 production from three nucleons. It is thus not clear whether the coalescence model for helium-3 production based on nucleons at kinetic freeze-out is also  a good approximation to the transport model calculations. Also, it has been found in Ref.~\cite{Song:2013qma} that baryon-antibaryon annihilation is more significant in Pb + Pb collisions at $\sqrt{s_{NN}}=$ 2.76 TeV than in Au + Au collisions at $\sqrt{s_{NN}}=$ 200 GeV, due to the longer evolution time in the hadronic stage at higher collision energies, which may affect the time when helium-3 is produced from the coalescence of nucleons.   

%\newpage

\end{document}